\documentclass[letterpaper,conference,10pt]{IEEEtran}
\usepackage[cmex10]{amsmath}
\usepackage{array}
\usepackage[font=footnotesize]{caption}
\usepackage[font=footnotesize]{subfig}
\usepackage{url}
\usepackage{cite}
\usepackage[pdftex]{graphicx}

\graphicspath{{./figures/}}
\DeclareGraphicsExtensions{.pdf,.jpg,.png}

\newcolumntype{L}{>{\arraybackslash}m{6cm}}

\usepackage{tikz}
\newcommand\copyrighttext{%
  \footnotesize \textcopyright 2016 IEEE. Personal use of this material is permitted. Permission from IEEE must be obtained for all other uses, in any current or future media, including reprinting/republishing this material for advertising or promotional purposes, creating new collective works, for resale or redistribution to servers or lists, or reuse of any copyrighted component of this work in other works.}
\newcommand\copyrightnotice{%
\begin{tikzpicture}[remember picture,overlay]
\node[anchor=south,yshift=10pt] at (current page.south) {\fbox{\parbox{\dimexpr\textwidth-\fboxsep-\fboxrule\relax}{\copyrighttext}}};
\end{tikzpicture}%
}
\begin{document}
\title{Toward Smart Moving Target Defense for Linux Container Resiliency}
\author{
\IEEEauthorblockN{Mohamed Azab}
\IEEEauthorblockA{
ACIS, ECE Department\\
University of Florida\\
mazab@vt.edu}
\and
\IEEEauthorblockN{Bassem Mokhtar}
\IEEEauthorblockA{
EE Department\\
Alexandria University\\
bmokhtar@alexu.edu.eg}
\and
\IEEEauthorblockN{Amr S. Abed}
\IEEEauthorblockA{
ECE Department\\
Virginia Tech\\
amrabed@vt.edu}
\and
\IEEEauthorblockN{Mohamed Eltoweissy}
\IEEEauthorblockA{
CIS Department\\
Virginia Military Institute\\
eltoweissymy@vmi.edu}}
\maketitle
\begin{abstract}
This paper presents ESCAPE, an informed moving target defense mechanism for cloud containers. ESCAPE models the interaction between attackers and their target containers as a ``predator searching for a prey" search game. Live migration of Linux-containers (\textit{prey}) is used to avoid attacks (\textit{predator}) and failures. The entire process is guided by a novel host-based behavior-monitoring system that seamlessly monitors containers for indications of intrusions and attacks. To evaluate ESCAPE effectiveness, we simulated the attack avoidance process based on a mathematical model mimicking the prey-vs-predator search game. Simulation results show high container survival probabilities with minimal added overhead.
\end{abstract}
\copyrightnotice
\IEEEpeerreviewmaketitle
\section{Introuction}
\label{sec:intro}
Linux containers running in a commercial cloud environments share the same kernel with containers from other customers, which increases the attack surface compared to the case of VM-based virtualization. To protect Linux containers in a shared environment, service providers are expected to monitor the behavior of the containers running on their system for suspicious activity. Upon detection of a possible threat, the service provider should take an action to protect the guest containers. While the most straightforward action is to kill the misbehaving container and inform the owner of the detected anomaly, such action may not be cost effective especially for long running stateful applications.

A more cost-effective alternative~\cite{laadan2010linux} is to take a snapshot of the running application while in safe state. Upon attack detection, the system simply rolls back to the most recent safe state saved. One drawback with this approach is when the last saved safe state is a vulnerable state and/or the attack is persistent, in which case the container will go into a continuous loop of restores. To overcome such limitations, we propose a nature inspired approach that aims at changing the container execution environment in order to mislead a persistent attack. The system aims to equip the attacker target (prey) with the tools needed to ESCAPE from the attacker (predator), e.g. by moving the container to a random remote host.%

In this paper, we use live migration of cloud containers as a moving target defense (MTD)~\cite{azab2013} mechanism against host-based persistent attacks. The MTD mechanism is guided by a host-based intrusion detection system (HIDS)~\cite{abed122015}\cite{abed092015} that monitors operating containers to detect potential anomalies or misbehaviors. The HIDS learns the behavior of all the containers running on the host, and upon detection of a change of behavior of one or more container, the HIDS signals the MTD module of the system to ESCAPE the affected container.
We considered two types of applications, stateless and stateful applications.%

The system reacts gradually and according to the application type. For stateful applications, the system is configured to start by using the checkpoint/restore mechanism before switching to the live-migration solution for a more persistent attack. Delaying the use of live-migration until the attack is known to be persistent (e.g. after N rollback attempts) saves the running application the overhead associated with live migration as compared to checkpoint/restore for non-persistent attacks.
For a stateless application, e.g. static web servers, upon attack detection, the system simply restarts the server or migrates the container to a new host while rerouting the associated network connections.  

The entire process is designed to mimic the famous ``Prey-vs-Predator" search game~\cite{alpern2006theory}. This game describes a scenario were multiple predators are targeting specific prey moving in a forest. We adopted this model into a guidance mechanism to guide the prey in its mission to evade the predator. In our scenario, the cloud is the forest. Each host in the cloud is a potential escape location for the prey. The Prey is the attacker (``predator") targeted container-encapsulated application.%

We built a simulation model to evaluate the effectiveness of the proposed approach in evading attacks. Additionally, we conducted preliminary experiments on our local ACIS cloud to evaluate the overhead of the live migration solution. Results showed the effectiveness of our approach with a limited to no overhead.%
%
ESCAPE was built to be as general as possible with minimal application customization. The main advantage of presenting generic defense tools is to give the user/system administrator the chance to select the most appropriate tools and applications that suits their needs with no constraints or limitations. 
ESCAPE supports general purpose Linux containers as a lightweight operating system virtualization technology. 
We are using Docker due to its popularity, small footprint, and fast instantiation. However, the same concepts discussed here can be applied to other Linux container systems, such as LXC and OpenVZ.
\section{System Overview}
\label{sec:overview}
ESCAPE mainly consists of three modules, namely the behavior monitoring module, the checkpoint/restore module, and the live migration module.
\subsection{Behavior Monitoring}
For monitoring purposes, we adopted the intrusion detection system introduced in~\cite{abed092015} to monitor the operating containers. The system is specifically tailored for monitoring the behavior of Linux containers, and depends on the fact that Linux containers communicates with the host kernel and the outer world by using system calls issued to the host kernel.

%
The monitoring system employs a background service running on the host kernel to monitor system calls between any Docker containers and the host Kernel. Upon start of a container, the service uses the open-source \texttt{sysdig} tool to trace all system calls issued by the container to the host kernel. The full trace is written to a log file that is processed in real time and used to learn the container behavior.

The monitoring system uses a bag of system calls (BoSC)~\cite{fuller2005} technique for learning the container behavior. 
The system reads the behavior log file epoch by epoch. For each epoch, a sliding window of size 10 is moved over the system calls of the current epoch, counting the number of occurrences of each distinct system call in the current window, and producing a BoSC. When a new occurrence of a system call is encountered, the corresponding index of the BoSC is updated. If the current BoSC already exists in the normal-behavior database, its frequency is incremented by 1. Otherwise, the new BoSC is added to the database with initial frequency of 1.

For detection mode, the system reads the behavior file epoch by epoch. For each epoch, a sliding window is similarly used to check if the current BoSC is present in the database of normal behavior database. If a BoSC is not present in the database, a mismatch is declared. The trace is declared anomalous if the number of mismatches within one epoch exceeds a certain threshold.
Upon attack detection, the resolution mechanism is called and the migration process is triggered to mimic the simulation scenario described in section~\ref{sec:math}.
\subsection{Checkpoint and Restore}
The application running inside the container, whether stateless or stateful, uses the host memory to host all runtime related libraries, calculations, and other volatile contents. It is too hard to recover these memory contents upon failure or migration events. ESCAPE were built to serve both types of applications with minor to no interference from the administrator or the programmer~\cite{azab2016}. Our primary goal is to avoid any container customization. We leveraged the encapsulated state of the application and used CRIU to dump the container memory into persistent set of files easy to share and recover. CRIU is used only on state-full applications based containers, we prefer not to use it on stateless type as the memory content and the executed states are not important for container restoration.

To enable checkpointing of running application, we are using an experimental version of Docker that integrates a checkpointing tool named CRIU to momentarily freeze the running container and its enclosed applications taking a live snapshot of the memory content and any used files. The dumped images are stored into persistent storage.

CRIU is a tool to checkpoint/restore running tasks in user space. CRIU momentarily freezes the running (container) runC process and all its sub-processes (user apps) and checkpoint it to a collection of image files that can be used to restore the container to the exact state later. Between these dump events, containers uses the host memory to operate to maximize the application response rate.

For a failed container to be restored, the hosting server must have access to the container image files, and the memory dump files. The container image files is usually large in terms of space. In our experiments, containers with full database server can be as large as 500MB. However, the memory dump is usually less than 10MB. The migration process for stateless type is much easier, we replicate the container on the destination server, then make a quick network switching between the source and destination. The replication process and container instantiation time is totally negligible as the original container will still be running (assuming that we are not recovering from failure), the migration process is entirely logical as the network connections are the only this that is going to ESCAPE  . ESCAPE adjust the ARP table for the NAT to point to the new server instead of the old one.%

For faster instantiation, quick recovery, and easy container migration, ESCAPE uses a remote shared storage as a container repository to store runC containers. Running the container from a remote storage gives instant access to multiple remote servers to instantiate such containers. Using remote repositories to host the base image of the container, massively reduced the time needed to move all the files between hosts in case of failure or live-migration. The only files that has to be synchronized between the source and destination servers are the memory dump which are so small and synchronize momentarily.%
%
\subsection{Live Migration}
%
For stateless applications, there is no need for checkpointing as the memory content is irrelevant for application re-execution. The relaxation of that constraint made it much easier for ESCAPE to recover such container in case of failure/attack.

Prior to starting the migration process, ESCAPE instantiates/selects a suitable destination. The current implementation follows the prey-vs predator model described in section~\ref{sec:math} to select the next destination for the migrating container. ESCAPE can also select a destination with far logical distance from the source to evade attackers. The logical distance is defined based on a heterogeneity scale. The more different the destination is in terms of configuration, network, and datacenter association the more likely for ESCAPE to select. ESCAPE aims to select such host to complicate the reapplication of the same cause of failure. However, considering that level of heterogeneity might induce resurrection conflicts.%

Upon selecting an appropriate destination, ESCAPE mounts the shared storage drive that holds the running container, and adjust the network settings to facilitate network relocation. 
The migration process then starts by checkpointing the container, killing the process on the original host, make an ARP update to change the MAC/IP assignment of the old server network interface to match the new one while mainlining the IP value, and restore the container and all enclosed applications on the destination server. The entire process occurs in matter of milliseconds. Following the aforementioned process guarantee almost zero downtime unless the source host fails completely during the process.
\section{Mathematical Model}
\label{sec:math}
\subsection{Model Design}
We have generated a mathematical model for evaluating the efficiency of the live migration mechanism of application containers adopted by ESCAPE to evading host-based cyber attacks. The developed model depends on theories and models of evolutionary search games~\cite{hofbauer1998evolutionary}.  We adopted the predator-prey model as a reference model to develop our mathematical model assuming an evasive prey ~\cite{oshanin2009survival}. Our model simulates and evaluates ESCAPE in mobilizing targeted containers across a set of attack-prone hosts within a virtual network considering some of containers running on those hosts are under attack.

We target in our ESCAPE performance evaluation study the survival probability of a targeted application container implemented on a host when operating over a cloud of networking hosts. 
The survival probability $P_{static}(t)$ of a static application container in a host is given by:
\begin{equation*}
\label{eq_1}
P_{static}(t)=e^{-\rho S(t)}=e^{-\frac{N}{V}S(t)}
\end{equation*}
Where
$\rho=N/V$ is the hacked host density, $N$ is the number of hacked hosts, $V$ is the number of all hosts in the studied network, and 
$S(t)$ is the mean number of distinct hosts visited by mobile attackers up to time $t$ and hosts' application containers malfunction. 
As $t$ increases, number of visited hosts increases (referring to the mobility of attacks and the possibility of having attacks widely spreading at many hosts). $S(t) \approx \pi t/\big(ln(t)\big)$ for a two-dimensional square area~\cite{oshanin2009survival}.

The survival probability $P_{mobile}(t)$ of a mobile application container is given by:
\begin{equation*}
\label{eq_2}
ln\Big(P_{mobile}(t)\Big)\approx {\Big(\frac{N}{V}\Big)}^2 ln\Big(P_{static}(T)\Big)
\end{equation*}

We assume that not all visited hosts by attackers will misbehave and only some related application containers will breakdown. In other words, we propose that a fraction of $S(t)$ will exist with exponential probability $1-e^{-t_d}$ where $t_d$ is the required time of detecting attacker's visits to a host. The mean number of hosts visited by mobile attackers that were successfully able to cause targeted containers to malfunction without detection, $S(t)_{success}$, is given by. 
\begin{equation*}
\label{eq_3}
S(t)_{success}=(1-e^{-t_d})S(t)
\end{equation*}

We consider two different growth models of attacks which are exponential and logistic growth as discussed in~\cite{alpern2006theory}. Those models discuss different rates of growth which refer to the spreading rate of attack in a networking context and how ESCAPE can succeed in mitigating such growth.
\begin{itemize}
\item Rapid increase of hacked hosts
\begin{equation*}
N(t)=N(0)e^{kt}
\end{equation*}
where $N(0)$ is the initial number of hacked hosts at time 0, and $k$ is a constant related to the increasing rate at any time $t$,
\item	Slow increase of hacked hosts
\begin{equation*}
N(t)=\frac{\mu N(0)e^kt}{N(0)e^kt+\mu-N(0)}
\end{equation*}
where $\mu$ is the carrying capacity which controls the increasing rate of $N$ that when $N$ approaches $\mu$, the increasing rate approaches zero. Also, $N(t)$ approaches $N(0)$ when $t \rightarrow 0$, and approaches $\mu$ when $t \rightarrow \infty$.
\end{itemize}
\subsection{Evaluation Study}
We conducted a simple simulation scenario of container-based cloud established by a set of connected hosts
under the following two assumptions: 1) We have one prey, $N$ hacked hosts, and $V$ total hosts (where $N\leq V$), and 2) Neighboring hosts to a hacked host are susceptible to the same attack.

The $N$ attackers are targeting the host of the valuable application container. When the attacker reaches a certain host, it spends some time on it after which all application containers working on that host fail. We considered both static and mobile container scenarios. Static containers are settled in one host, while mobile containers can migrate from a host to another one.
We study the effect of having small and large number of hacked hosts ($N$) on the survival probabilities considering two cases:
\begin{enumerate}
\item \emph{Static $N$}: 
$N$ is fixed (${dN(t)}/{dt}=0$) during the simulation, i.e. the attackers move from one host to another preserving the total number of hacked hosts ($N$).
\item \emph{Dynamic  $N$}: 
$N$ varies with the simulation time ($dN(t)/{dt}\neq0$), i.e., attackers migrate to other hosts and attack new hosts (number of hacked hosts increases with the simulation time).
We repeat the scenarios when N has rapid and slow increasing rates and how N affects the survival probabilities of targeted containers
\end{enumerate}
We then repeat the previous scenarios when applying the concept of attacker success on survival probabilities of targeted application containers.

\begin{figure}[htbp]
\begin{center}
\includegraphics[height=2.25in, width=0.5\textwidth]{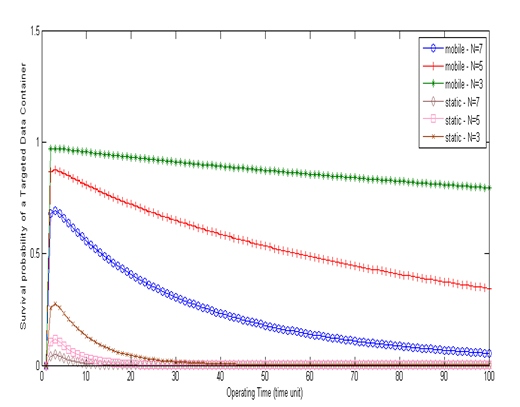}
\caption{Survival probability of a static/mobile container at different number of hacked hosts ($N$)}
\label{fig_5}
\end{center}
\end{figure}

\begin{figure}[htbp]
\begin{center}
\includegraphics[height=2.25in, width=0.5\textwidth]{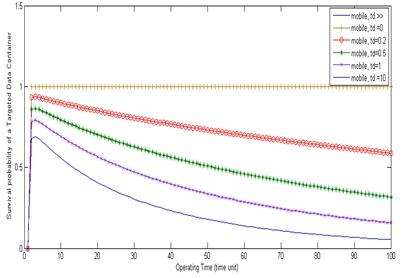}
\caption{Survival probability of a targeted mobile container at 7 hacked hosts and various attack detection times}
\label{fig_6}
\end{center}
\end{figure}

\begin{figure}[htbp]
\begin{center}
\includegraphics[height=2.25in, width=0.5\textwidth]{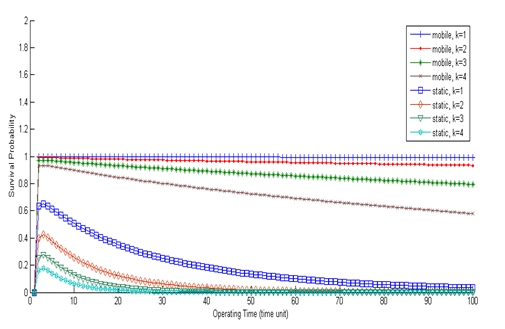}
\caption{Survival probability of a static/mobile targeted container at various increasing rates of logistic attack growth}
\label{fig_8}
\end{center}
\end{figure}

\begin{figure}[htbp]
\begin{center}
\includegraphics[height=2.25in, width=0.5\textwidth]{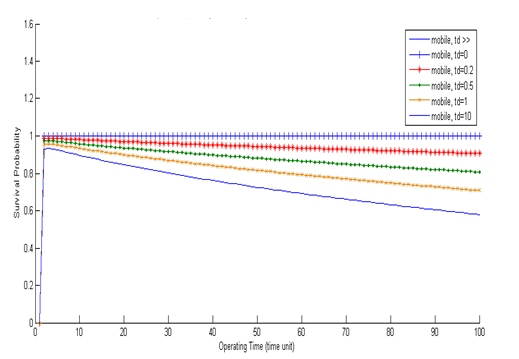}
\caption{Survival probability of a mobile targeted container at logistic attack growth with various attack detection time}
\label{fig_9}
\end{center}
\end{figure}

Figure~\ref{fig_5} shows the impact of changing the number of hacked hosts ($N$) on a network of 20 hosts where the targeted contained located in one host. The figure shows that we can have higher survival probabilities in case of having mobile targeted application container capability.
Figure~\ref{fig_6} shows that at various detection times and 7 hacked hosts, we can get better survival probabilities at smaller detection times. If we compared the obtained results in figure~\ref{fig_6} with the simulated survival probability value at case $N=7$ at figure~\ref{fig_5}, we can notice the improvement in the results at small detection times.
Figures~\ref{fig_8} and ~\ref{fig_9} show the impact of having logistic growth of attacks on the survival probability of a static/mobile application containers in case of having various increasing rates and attack detection times, respectively. As we have small attack detection times, we can mitigate the high growth rate of attacks and their spread in many hosts.
\section{Conclusion}
\label{sec:conclusion}
In this paper, we introduced ESCAPE, a novel nature-inspired mechanism for attack detection and avoidance. ESCAPE deploys a novel interrelated real-time monitoring and live migration of cloud containers. 
We proposed a mathematical model based on evolutionary search games to guide ESCAPE. The same model is used to simulate ESCAPE operation and to evaluate its performance.  
ESCAPE was able to effectively and efficiently detect and avoid persistent attacks achieving high survival probabilities of legitimate application containers.%
\bibliography{references}
\end{document}